\documentclass[aps,twocolumn,prd,nofootinbib]{revtex4}
\usepackage{amssymb}
\usepackage{graphicx}
\usepackage{soul}
\newcommand{\be}{\begin{equation}}
\newcommand{\ee}{\end{equation}}
\newcommand{\bea}{\begin{eqnarray}}
\newcommand{\eea}{\end{eqnarray}}
\newcommand{\mpl}{m_{\rm pl}}

\newcommand{\fnl}{f_{\rm NL}}
\newcommand{\ns}{n_{ \rm s}}

\begin{document}



\title{Non-Gaussianity from the hybrid potential}
\author{David Mulryne$^{1,2}$, Stefano Orani$^2$ and Arttu Rajantie$^2$}

\affiliation{$^{1}$Astronomy Unit, School of Mathematical Sciences, 
Queen Mary, University of London,  
London, E1 3NS. \\$^2$Department of Physics, Imperial College, London, SW7 2AZ.  }
\date{21 July 2011}
\begin{abstract}
We study the hybrid inflationary potential in a regime where the 
defect field is light, and more than 
60 e-folds of accelerated expansion occur after the symmetry breaking transition. 
Using analytic and numerical techniques,
we then identify parameter values within this regime for which  
the statistics of the primordial curvature perturbation are significantly 
non-Gaussian.  Focusing on this range of parameters, we provide 
a specific example which leads to an observationally 
consistent power spectrum, and a level of non-Gaussianity 
within current WMAP bounds and in 
reach of the Planck satellite. An interesting feature 
of this example is that the initial conditions at horizon crossing 
appear quite natural.
\end{abstract}
\maketitle

\section{Introduction}
                                                            
An inflationary epoch is the most promising  
model for the origin of large-scale structure in the 
universe \cite{inflation} (for reviews see \cite{reviews}), 
naturally accounting  
for current observations which indicate that the universe's 
structure originated from  
near-scale invariant and almost Gaussian 
fluctuations \cite{observations,wmap}. 
The small 
deviations from scale-invariance and Gaussianity are 
key probes of inflation, 
allowing competing models to be distinguished 
from one another by present
and future observations.

In canonical, single field inflation, the curvature 
perturbation produced at horizon crossing is 
extremely close to 
Gaussian, and is conserved
through the entirety 
of the subsequent evolution \cite{Maldacena:2002vr,Bardeen,Lyth,zetaConv}. 
On the other hand,  when more than one 
field is light at horizon 
crossing, isocurvature modes are produced, which later 
can source an additional contribution to the 
curvature perturbation. 
This can cause the curvature 
perturbation to become sufficiently 
non-Gaussian for this signal  
to be detected by future observations, 
or the model  
constrained by present ones.  
The current WMAP constraints on the $\fnl$ 
parameter, which measures the deviation of the three--point 
correlation function from zero, is $-10<\fnl <74$ at $95\%$ confidence 
level, but $\fnl \approx 32\pm 21$ at the $68\%$ level~\cite{wmap}. Though premature, 
it is tempting to 
consider these numbers as a `tentative hint' of 
a non-Gaussian signal. The recently launched PLANCK satellite, will 
help resolve the issue. In the absence of a detection it is expected 
to give the constraint $|\fnl|<5$.

With these issues in mind, 
considerable effort has been invested in 
developing methods to calculate how 
non-Gaussianity evolves on super-horizon scales during inflation, 
\cite{Lyth:2005fi,Vernizzi:2006ve, otherMethods},
into identifying the features 
an inflationary potential must 
possess and the initial conditions required, 
to lead to a `large' non-Gaussianity ($\fnl>1$) \cite{largeNG1,largeNG2,Kim:2010ud,Elliston:2011dr}. 
Many scenarios have also been studied in which the generation 
of non-Gaussianity occurs due to post inflationary effects, 
such as in curvaton \cite{curv}, preheating~\cite{preh}, modulated (p)reheating \cite{mod}, 
and models with an inhomogenious end of inflation \cite{inHom1,inhom2} 
(see Ref.~\cite{Alabidi:2010ba} for 
connections between these three scenarios).

Of particular interest are effects which 
occur for inflationary potentials which are 
well motivated 
from a particle physics perspective.  
The hybrid potential is one such example, 
and moreover is one of the most widely studied potentials 
in the literature. 
First introduced as an inflationary model by 
Linde \cite{Linde:1993cn}, 
it is a two-field potential of the form
\be
V(\phi,\chi)=\frac{1}{2}m^2\phi^2+\frac{1}{2}g^2\phi^2\chi^2+\frac{\lambda}{4}(\chi^2-v^2)^2\,.
\label{eq:hybridPot}
\ee
As a quantum field theory, this is attractive because 
it is renormalisable and does not require unnatural cancellation 
of radiative corrections. 
In fact, potentials of this type are common in supersymmetric models \cite{SUSY}.

In this paper we will investigate the hybrid potential (\ref{eq:hybridPot}) using a different choice of parameters from the original proposal. We find that this leads to a viable cosmological model, with perturbations that are compatible with current observations and
level of non-Gaussianity that would be observable with the Planck satellite.

\section{Original Hybrid Inflation}
In the original version of the scenario \cite{Linde:1993cn} the 
$\phi$ field 
plays the role of the inflaton. 
Initially, $\phi$ has a value larger than the 
`critical' value, $\phi>\phi_{\rm crit} \equiv \sqrt{\lambda} v/g$, 
but still below the Planck scale $\phi\ll \mpl=(8\pi G)^{-1/2}$. The 
latter requires $\lambda v^2<g^2 \mpl^2$.
The field $\phi$ then rolls down the potential towards smaller 
values until it reaches $\phi_{\rm crit}$, at which 
point $\chi$ becomes non-zero and inflation ends.

We can generally consider a multi-field model with scalar 
fields $\varphi^\alpha$ labelled by an integer $\alpha$. In the case 
of the hybrid model, the fields are $\varphi_1=\phi$ 
and $\varphi_2=\chi$. Defining the slow-roll parameters,
\be
\epsilon_\alpha= \frac{\mpl^2}{2}  \left (\frac{V_{,\alpha }}{ V} \right )^2\,,
\label{eq:epsilon}
\ee
and
\be
\eta_{\alpha\beta} = \mpl^2 \frac{V_{,\alpha \beta}}{V}\,,
\label{eq:eta}
\ee
the universe inflates when $\epsilon \equiv \sum_{\alpha} \epsilon_{\alpha} < 1$ 
and when $\epsilon\ll1$
the fields satisfy the `slow-roll' equations of motion 
\be
\frac{d \varphi^{\alpha} }{d N} = -\sqrt{2\epsilon_\alpha} \mpl\,.
\label{eq:sr}
\ee

The standard hybrid model assumes that the potential (\ref{eq:hybridPot}) is 
dominated by the constant term during inflation, 
and this requires $\lambda v^4>m^2\mpl^2$.
The parameters are also chosen such that when $\phi$ reaches 
$\phi_{\rm crit}$, $\chi$ rapidly grows 
and reaches the global minimum at 
$\chi = \pm v$ in less than an e-fold \cite{Copeland:2002ku}, 
thus ending inflation.  
This requires that the potential is steep 
enough in the $\chi$ direction to avoid 
further inflation, or more precisely $v<\mpl$.
Also, $\phi$ must pass through $\phi_{\rm crit}$ fast enough that
$\chi$ remains light for less than an e-fold.
Utilising Eq. (\ref{eq:sr}) and (\ref{eq:epsilon}) and linearising  
about $\phi_{\rm crit}$ such that $\Delta \phi = \phi - \phi_{\rm crit}$, 
$\Delta N = \sqrt{\epsilon_{\phi_{\rm crit}}} \Delta \phi$, and  
$m^2_\chi = 2 g^2 \phi_{\rm crit} \Delta \phi $. One finds that $\chi$ 
is light for $N \approx \lambda v^6/(m^2 \mpl^4)$, and hence this combination 
of parameters must be less than unity.
To summarise, therefore, the original scenario requires the parameter choices
\begin{eqnarray}
\label{eq:originalParameters}
\lambda v^2 &<& g^2 \mpl^2,\nonumber\\
\lambda v^4 &>& m^2 \mpl^2,\nonumber\\
v&<&\mpl,\nonumber\\
\lambda v^6 &<& m^2 \mpl^4.
\end{eqnarray}

It is important to realise, however, that these constraints 
are not all necessarily required for a viable 
cosmological model. They simply correspond to the 
specific physical scenario assumed in the original papers.
Despite having many appealing features, this original scenario suffers 
from two important issues. First, 
the phase transition generates a network of
domains where different choices of vacuum have been made. 
The domains are separated by domain 
walls, which are  incompatible with the observable 
universe. This issue can be 
overcome by using a multi-component $\chi$ field, 
and possibly coupling it to a gauge field to 
eliminate Goldstone bosons, but this complicates 
the model significantly. Second, the scenario
predicts a blue spectral index ($n_s>1$). This is 
because for inflation driven solely by the $\phi$ field, 
$\ns = 1 - 6 \epsilon^*_\phi+2 \eta^*_{\phi \phi}$ (where $^*$ labels 
the time at which observational wavelengths became greater than the 
cosmological horizon roughly $60$ e-folds before the end of inflation), 
and for the parameter choices discussed above, 
$\eta^*_{\phi \phi}$ is positive and satisfies 
$\eta_{\phi \phi}^*\gg\epsilon^*_\phi$. This is in conflict 
with WMAP and baryon acoustic 
oscillations measurements which give  $n_s = 0.968 \pm 0.012$ 
with $68\%$ confidence level \cite{wmap}. 

\section{Hybrid inflation with two light fields} 
Given the motivation for the form of potential (\ref{eq:hybridPot}), 
it is interesting to study its consequences outside of the 
parameter ranges (\ref{eq:originalParameters}) which 
lead to the original scenario. For example, considerable 
interest has been devoted to parameter choices which 
make both fields light (see for example 
\cite{Randall:1996ip, GarciaBellido:1996qt}), 
and there have been a number of investigations into 
the consequences of this situation, such 
as the resultant non-Gaussianity \cite{barnaby}.  
If the waterfall field is light at the 
symmetry breaking transition for more than an e-fold, however, 
the amplitude of 
density perturbations on scales associated with this transition is  
in tension with primordial black hole 
constraints \cite{GarciaBellido:1996qt}.  

Recently, it has been noted by Clesse \cite{Clesse:2010iz} that 
this conclusion does not apply 
if the waterfall field is light but more than $60$ 
e-folds of inflation occur after the 
phase transition. This is because any black holes 
produced are diluted by the expansion, and 
moreover, this also solves the defect problem, as  
defects are pushed onto scales 
larger than the present day observable universe. 
Clesse considered a parameter range identical to the 
original hybrid scenario, except that he choose parameters 
such that $\phi$ stays sufficiently close to $\phi_{\rm crit}$ for $\chi$ to 
be light for more than $60$ e-folds, requiring  
$\lambda v^6 \gg 60 m^2 \mpl^2$.
In this minimal adjustment to the original scenario, both fields still move 
sub-Planckian distances, but the velocity of the $\chi$ field is 
much greater than the $\phi$ field for the entire 
evolution. This implies\footnote{This is an 
approximation because in reality the dynamics of 
both fields 
contribute to the spectral index, as we shall 
see in the next section.}
that 
$n_s = 1 - 6 \epsilon_\chi+2 \eta_{\chi \chi}$
and because the model is still vacuum dominated and 
$\eta_\chi$ is negative (because inflation 
occurs after the transition where $\chi$ 
has a negative mass), a red spectral index results. 
The final issue with the original scenario is therefore resolved.

Clesse's scenario does not lead to observable levels of non-Gaussianity. 
In this paper we
show that there are other parameter choices which lead 
to perturbations that are compatible with observations but 
are significantly non-Gaussian.

\section{The  $\mathbf {\delta}$N formalism.} 
\label{sec:deltaN}
To calculate the level of non-Gaussianity as well as the 
amplitude of the power spectrum and the spectral index, 
we will employ the $\delta N$ formalism\cite{Starobinsky:1986fxa, Sasaki:1995aw, Lyth:2005fi},  
implemented with numerical and analytic techniques.

The $\delta N$ formalism, assumes the separate 
universe approximation, in which points in the 
universe separated by more than a horizon size 
evolve like a `separate universe', the dynamics 
following from the local energy density and pressure, 
and obeying the field equations of a Friedmann-Robertson-Walker (FRW) 
universe (see for example \cite{Lyth, Wands:2000dp}). The $\delta N$ formalism then notes that 
the uniform density curvature perturbation 
between two such points, $\zeta$, is given by the difference in 
e-folds of expansion between them, from some initial 
shared flat hypersurface labelled $*$, to some final 
shared hypersurface of uniform density labelled $f$,
\be
\zeta = \delta N^f_*.
\ee 
We use $*$ to label the flat hypersurface, since 
in order to make observational predictions for an inflationary 
model, this surface must be 
defined at the time when obervable scales exited the cosmological 
horizon. Moreover, 
the final hypersurface $f$ must be taken at some much 
later time when the dynamics are adiabatic
and $\zeta$ is conserved.
This can happen, for example, long after inflation ends, 
when the dynamics are 
dominated by a single fluid.

When slow-roll is a good approximation at horizon exit, $\zeta$ 
is completely determined by
the field perturbations $\delta\varphi^\alpha_*$ at that time,
\be
\zeta=\delta N_*^f(\delta\varphi^\alpha_*).
\ee
The field perturbations are extremely close to 
Gaussian \cite{Gaussian}, and if their amplitude 
is sufficiently small, the statistics of 
the curvature perturbation can be determined in a 
simple manner by Taylor expanding
\be
\delta N^f_* \approx \frac{\partial N^f_* (\phi_*, \chi_*)}{\partial \varphi^{\alpha}_*} \delta\varphi^{\alpha}_* + \frac{1}{2}\frac{\partial^2 N^f_*}{\partial \varphi^{\alpha}_*\varphi^{\beta}_*} \delta\varphi^{\alpha}_*\delta\varphi^{\beta}_*\,,
\label{eq2}
\ee
where here and from here on we employ the summation convention, and subsequently we 
will employ the notation in common use, $N_{\alpha}={\partial N^f_* (\phi_*, \chi_*)}/{\partial \varphi^{\alpha}_*}$.

One then finds that the amplitude of the power spectrum
is given by 
\be
A_\zeta^2 = N_{\alpha}N_{\alpha}\frac{H_*^2}{4\pi^2}\,,
\ee 
the spectral index $n_s$  
by \cite{Sasaki:1995aw} 
\be
n_s =1   - 2\epsilon_* + \frac{2}{H_*}\frac{\dot{\varphi}^{\alpha}_*N_{\beta}N_{\alpha\beta}}{N_{\alpha}N_{\alpha}} \,,
\label{eq:spectrum}
\ee
which generalises the single field expression we employed earlier, 
and the amplitude of the reduced bispectrum by \cite{Lyth:2005fi}
\be
 \fnl = \frac{5}{6}\frac{N_{\alpha}N_{\beta}N_{\alpha\beta}}{(N_{\alpha}N_{\alpha})^2}\,.
 \label{eq6}
 \ee

\section{Analytic estimates}
\label{sec:AG}

In the end we will analyse the model using numerical methods, but to identify suitable parameter ranges, let us first use analytic approximations to model the behaviour of the model.
One particular case in which analytic calculations are possible is when the potential is
separable,
$$V(\phi,\chi) = V_\phi(\phi) + V_\chi(\chi).$$
Of course, the hybrid potential Eq.~(\ref{eq:hybridPot}) is not separable, but this should still be a reasonable approximation if the non-separable term $g^2\phi^2 \chi^2 $ is sub-dominant to the other terms. We will therefore use this approximation, and later check its validity 
using full numerical simulations.

For separable potentials, there exist analytic expressions 
for the derivatives of $N$ \cite{Vernizzi:2006ve, GarciaBellido:1995qq}. 
The starting point for these formulae are the slow-roll equations 
of motion Eq. (\ref{eq:sr}).
It follows that 
\be
N = \int^f_* d \phi \frac{-V(\phi, \chi) }{  \mpl^2 V_{,\phi} } =  \int^f_*d \phi  \frac{-V_\phi}{ \mpl^2 V_{,\phi} }+ \int^f_* d \chi \frac{-V_\chi }{ \mpl^2 V_{,\chi} }\,.
\ee
Therefore, for example, 
\be
N_{,\chi} =\left. \frac{V_\chi}{ \mpl^2 V_{,\chi} }\right |_* - \left. \frac{V_\chi}{ \mpl^2 V_{, \chi} }\right |_* \frac{\partial \phi^f}{\partial \phi_*}\,.
\ee
The complexity of the calculation then reduces to determining 
the partial derivative of the final field 
value on a constant energy density hypersurface with respect 
to the initial field value. However, when the dynamics become 
adiabatic during inflation, this derivative tends to zero. Therefore, 
one usually finds that\footnote{It is possible that this 
is not the case if $V_\chi/V_{,\chi}|_f$ diverges as the derivative 
tends to zero (see \cite{Elliston:2011dr} for a full discussion).}
 $N _{\chi}= V_\chi/(\mpl^2 V_{,\chi}|_*)$.
The second derivatives 
of $N$ then follow by simple differentiation.

One known mechanism which gives rise to a large non-Gaussianity 
in multiple field inflation is for one of the 
fields to be very close to a maximum of its 
self-interaction potential at the time 
of horizon crossing \cite{Kim:2010ud,Elliston:2011dr}, and for the magnitude 
of its tachyonic mass--squared to be much greater than 
the vacuum energy at this hilltop. In this region the $\chi$ part of 
the potential can 
be approximated by $V_\chi = V_0 - 1/2 m_\chi^2 \chi^2$, 
and the condition required is that $m_\chi^2 \mpl^2  \gg V_0$ \cite{Kim:2010ud}. 
Of course the field must also be light at horizon crossing, 
and hence $m_\chi^2 < H^2$, but this condition can be simultaneously 
satisfied assuming the potential is not vacuum dominated, 
and that another field contributes significantly to 
the total energy density.

Following Ref.~\cite{Kim:2010ud} 
we can understand this condition by considering that when $m_\chi^2 \chi^2 < V_0$, 
$N_{\chi} \approx -V_0/ (\mpl^2 m_\chi^2 \chi_*)$, which becomes large in the 
limit that $\chi_* \rightarrow 0$. In this limit, the 
contribution to $\delta N$ from the $\chi$ field likely 
dominates over all other contributions, 
and moreover  
$N_{\chi \chi} \approx V_0/ (\mpl^2 m_\chi^2 \chi_*^2)$, 
and $\fnl \approx (5 \mpl^2 m_\chi^2 )/(6 V_0)$. 

For the hybrid potential well after the symmetry 
breaking transition, the shape of the potential is exactly of the form 
discussed above, with the non-separable term sub-dominant. In that case, 
$V_0 = \lambda v^4/4$,  
$m_\chi^2 =  \lambda v^2$ and hence for this regime we expect 
\be
\fnl \approx \frac{10}{3} \frac{ \mpl^2}{v^2 }\,,
\label{anpred}
\ee
assuming that $\chi_*$ is 
sufficiently small.

In principle, therefore, 
for $v<\mpl$ it is possible for this hybrid potential to give rise to a 
large positive non-Gaussianity.   The additional constraint 
that the $\chi$ field be light, 
however, introduces the restriction that 
$1/2 m_\phi^2 \phi_*^2 \gg V_0 = \lambda v^4/4$, and 
the $\phi$ part of the potential dominates the energy density. This 
implies we must move beyond the original 
parameter choices and consider large field values of $\phi$ 
in order that sufficient inflation occurs. This in turn 
implies that $\phi_{\rm crit} \geq 16 \mpl$, 
where equality leads to roughly $60$ e-folds after the transition. 
Hence $\phi* \approx 16 \mpl$.  
In this case $N_{\phi} = \phi_*/(2\mpl^2)$, and  
the condition that the $\chi$ field sources the dominant contribution to $\delta N$, 
$|N_{\chi}| > |N_{\phi}|$, implies that 
$v^2 > 32 \chi_* \mpl$.   
We 
arrive at the curious situation in which the 
$\phi$ field sources inflation, while the contribution 
due to the $\chi$ field dominates $\zeta$, and 
can be extremely non-Gaussian. This can be summarised by the 
following requirements on the parameters 
\begin{eqnarray}
\lambda v^2 &\gg& g^2 \mpl^2,\nonumber\\
  m^2 \phi^{*2} &\gg& \lambda v^4,\nonumber\\ 
  v&<&\mpl,\nonumber\\ 
  \phi_* &\approx& 16\mpl,\nonumber\\
\chi_* &\lesssim& 0.03 v^2/\mpl
\label{eq:ourParameters}
\end{eqnarray}
On the other hand, consideration of Eq.~(\ref{eq:spectrum}) implies that the 
spectral index is generically red tilted and close to scale 
invariant, as will be confirmed by the full numerical 
simulations which follow.

Guidance from analytic arguments, therefore, appears to lead to a 
large non-Gaussianity in the following scenario. 
Long before the last $60$ e-folds of inflation, $\phi$ begins its evolution 
with $\phi>\phi_{\rm crit}$, 
and some value of $\chi$. 
Since the mass of $\chi$ is initially positive it will evolve 
towards its minimum at $\chi=0$ as $\phi$ evolves towards $\phi_{\rm crit}$. 
Assuming the classical evolution of $\chi$ reaches $\chi=0$ before $\phi$ 
reaches $\phi_{\rm crit}$, then at this time the value of $\chi$ 
will be dominated by its quantum diffusion during the period 
before the transition for which it was light.  
After the transition for the parameter 
choices discussed, more than $60$ e-folds of 
inflation occur and $\chi$ is still light when 
$\phi =16 \mpl$ roughly $60$ e-folds 
before the end of inflation. Its
vev at this time will then likely still be close to $\chi=0$. 
The $\chi$ field will only roll significantly when 
$\chi$ becomes heavy. If this occurs 
before $\phi$ evolves to its minimum, it will roll during 
the slow-roll inflationary phase, otherwise $\phi$ will begin 
oscillating and begin to decay into radiation before $\chi$ 
begins to roll. In the later case $\chi$ will 
perhaps source an extremely short secondary inflationary 
phase as it rolls. It is clear, however, that Eq.~(\ref{anpred})
will remain 
at best a \emph{rough estimate}. This is why we 
employ numerical simulations in our study.

For completeness we note that when $\chi_*$ becomes larger and Eq.~(\ref{anpred}) becomes obsolete, it is still possible to find an analytic estimate for the asymptotic value of $\fnl$.
At $\chi_*\sim v^2/\mpl$, we reach a regime where $|N_\phi|>|N_\chi|$. If the numerator of Eq.~(\ref{eq6}) is still dominated by $\chi$, we find
\be
\fnl=\frac{5}{6}\frac{N_\chi^2N_{\chi\chi}}{N_\phi^4}\approx \frac{v^6}{\mpl^2\chi_*^4},
\label{equ:fnltrans}
\ee
indicating that $\fnl$ drops sharply. At even larger values of  $\chi_*$, we find
$N_{\phi\phi}>N_{\chi\chi}$, which implies  $\fnl = 5N_{\phi\phi}/(6N_\phi^2)$.
Using $N_\phi=\phi_*/2\mpl^2$ and $N_{\phi\phi}=1/2\mpl^2$, we obtain
\begin{equation}
\fnl = \frac{5\mpl^2}{3\phi_*^2}\approx 0.007.
\label{anpredphi}
\end{equation}
Hence, we should find a smooth but clear transition from the significant non-Gaussianity predicted by Eq.~(\ref{anpred}) to practically Gaussian single-field behaviour (\ref{anpredphi}) at around $\chi_*\approx 0.03v^2/\mpl$.

\section{Initial conditions}
\label{sec:ics}

The scenario discussed above requires a small initial value for 
the waterfall field, $\chi_*\ll v^2/\mpl$, at $\phi=\phi_{\rm crit}$.
In this section we discuss the naturalness of this condition.
This is an instance of the `measure problem' which is unresolved in general. 
However, we can apply some heuristic arguments. 

In Section \ref{sec:AG}, we outlined a plausible sequence of events 
in which, long before the hybrid transition, both $\phi$ and 
$\chi$ took large values, and the classical evolution 
of $\chi$ tended to zero before $\phi$ reached $\phi_{\rm crit}$. 
The initial values of $\phi$ and $\chi$ can be motivated by 
assuming a phase of eternal inflation during which the path the 
fields follow is dominated by quantum fluctuations, rather than 
classical rolling. The condition for such behaviour is that the 
distance moved in field space in a Hubble time, is less than  
a typical quantum fluctuation \cite{eternal}
\be
\sqrt{\dot{\phi}^2 +\dot{\chi}^2}> \frac{H^2}{\sqrt{2 \pi}}\,.
\label{eq:eternal}
\ee
Using the slow roll equations of motion (\ref{eq:sr}) this becomes a constraint 
on the field values, and defines a one dimensional surface in field space at 
which the dynamics becomes predominantly classical.

Beginning at this surface, therefore, 
one might wonder whether $\chi$ will indeed reach $\chi=0$ (according to its classical 
rolling) before  $\phi$ reaches $\phi_{\rm crit}$. The answer to this question 
will be extremely parameter dependent, and will likely also depend on the 
position on the surface from which the fields originate. It is easy, however,  
to build up a rough picture of the expected behaviour. 
Considering cases where both $\phi$ and 
$\chi$ are significantly displaced from zero initially, $\chi$ will likely 
initially be the more massive field for the parameter choices we are focusing on (\ref{eq:ourParameters}). 
This is because $\chi$ feels a mass of $1/2 \lambda \chi^2 +g^2 \phi^2$ in comparison with 
$m^2 + g^2 \chi^2$, the mass of $\phi$. Moreover, 
we expect $\lambda>g^2$, unless $v$ is much less than $\mpl$. 
In the following we will assume this condition on $\lambda$ and $g$.   
The mass of 
$\phi$ would only be greater initially, therefore, 
if $m^2 \gtrsim \lambda \chi_{\rm e}^2 + g^2 \phi_{\rm e}^2$ 
(where subscript $\rm e$ represents values on the eternal inflation surface). 

Assuming 
that $\chi$ is indeed more massive initially, it will evolve rapidly towards 
$\chi=0$, while $\phi$ will remain nearly frozen. 
As $\chi$ decreases, there will come a time when $\lambda \chi^2 < m^2$. At this point 
if the condition $m^2 \gtrsim g^2 \phi^2_{\rm e}$ is met then the 
$\chi$ field will cease to be the more massive, and 
its evolution will slow significantly, and it will likely not 
reach $\chi=0$. On the other hand, if this condition is not met we 
expect that $\chi$ does reach zero.  The necessary condition for 
$\chi$ to reach zero, therefore, is that $m^2 < g^2 \phi^2_{\rm e}$, which ensures 
that $\chi$ is always the more massive field. 

This condition is likely only very approximate, but from some limited 
numerical probing it appears to capture the rough behaviour of the fields.  
In Section~\ref{sec:workExample} we will look carefully 
at a particular parameter choice and probe the surface defined by Eq.~(\ref{eq:eternal}). We find 
a complicated picture, however this rough analysis is a useful aid.

\begin{figure}
\centering
\includegraphics[width= 8.6cm]{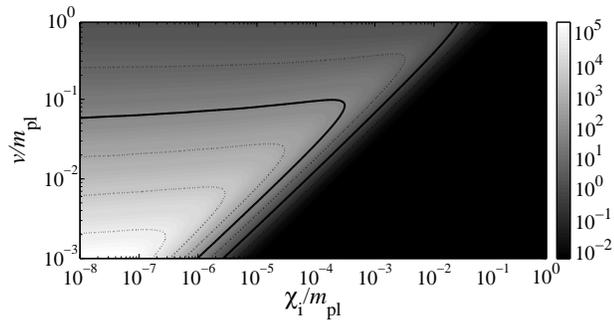}
\caption{The non-Gaussianity parameter $\fnl$ as a function of $\chi_i$ and $v$, with other parameters 
given by Eq.~(\ref{equ:paramchoices}). Lighter shades indicate higher $\fnl$. High 
values $\fnl\gtrsim 100$ are ruled out by WMAP, and small values $\fnl\lesssim 1$ are likely to be unobservable.}
\label{confnl}
\end{figure}
\begin{figure}
\centering
\includegraphics[width= 8.6cm]{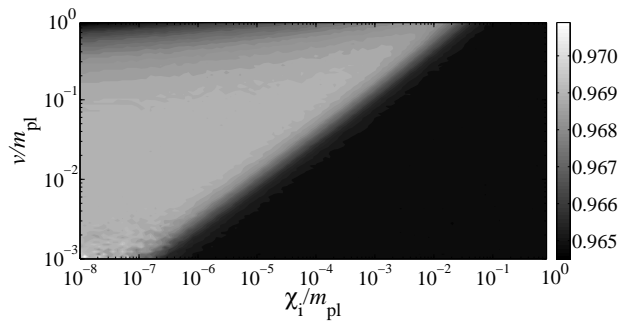}
\caption{The spectral index $n_s$ 
as a function of $\chi_i$ and $v$, with other parameters given by Eq.~(\ref{equ:paramchoices}). Lighter shades indicate higher $n_s$. All values obtained are compatible with WMAP observations.}
\label{conns}
\end{figure}
\begin{figure}
\centering
\includegraphics[width= 8.6cm]{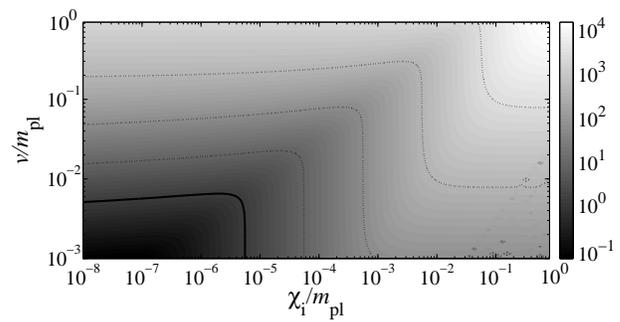}
\caption{The ratio $\chi_*/H_*$ 
as a function of $\chi_i$ and $v$, with other parameters given by Eq.~(\ref{equ:paramchoices}). Lighter shades indicate higher $\chi_*/H_*$. The thicker contour line corresponds to $\chi_*/H_*$=1. Values below this are unnatural because quantum fluctuations would generally give $\chi_*\gtrsim H_*$}
\label{conDH}
\end{figure}

\begin{figure}
\centering
\includegraphics[width= 8.6cm]{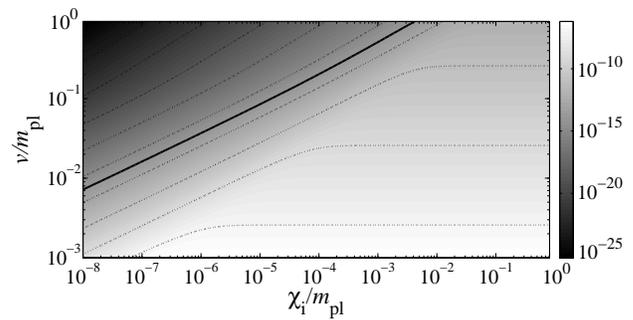}
\caption{The dimensionless parameters $\lambda$ 
as a function of $\chi_i$ and $v$, with other parameters given by Eq.~(\ref{equ:paramchoices}). Lighter shades indicate higher $\lambda$. The thicker contour line corresponds to $\lambda=10^{-14}$.}
\label{conLambda}
\end{figure}

\section{Numerical analysis}
\label{sec:numerical}
Having identified the interesting parameter range, we now calculate the
statistics of the perturbations numerically.
We evolve the full non-slow-roll equations 
of motion 
\begin{eqnarray}
\label{equ:eoms}
\ddot{\varphi}^\alpha + 3 H \dot{\varphi}^\alpha + \frac{\partial V}{\partial \varphi^\alpha} &=& -\Gamma_\alpha \dot{\varphi}_\alpha\,,\nonumber\\
\dot{\rho}_{\rm rad} + 4 H \rho &=& \Gamma_\alpha \dot{\varphi}_\alpha^2\,,
\end{eqnarray}
where
\be
H^2 = \frac{1}{3 \mpl^2}\left(\frac{1}{2}\dot{\varphi}^\alpha \dot{\varphi}^\alpha +V +\rho_{\rm rad}\right)\,,
\ee
and we have introduced a small decay term $\Gamma_\alpha$ to describe the decay of the scalar 
fields into radiation. As long as the decay rate is fairly slow, the produced perturbations are 
largely independent of it, so we do not attempt to use a realistic value but only check that our 
results do not depend on it.

For simplicity we reduce the number of parameters which 
can be varied, by fixing
\bea
\phi_{\rm crit} &=& \sqrt{1000}\mpl,\nonumber\\
m&=&\sqrt{\lambda}v,\nonumber\\
g&=& \sqrt{\lambda} v/ \phi_{\rm crit},\nonumber\\
\Gamma&=& 10^{-1} \sqrt{\lambda} v,
\label{equ:paramchoices}
\eea
and we explore the ranges
\bea
10^{-3}\mpl <&v&<\mpl,\nonumber\\ 
10^{-6}<&\chi_{\rm i}/v&<1.
\eea

Our parameter choice effectively means that the 
magnitude of $H_*$ and of $\Gamma$ is proportional to $\sqrt{\lambda}$, 
and so can be scaled simply by changing the 
value of $\lambda$. On the other hand, since the derivatives of $N^f_*$, which 
enter formulae for observational quantities, are unaltered by such a scaling, changing 
the value of $\lambda$ leaves these derivatives unaltered. 
For each 
choice of $\chi_{\rm i}$ and $v$, we fix $\lambda$ such that the 
resulting amplitude $A_\zeta$ is in agreement with the observational 
requirement. The rescaling properties just discussed, however, 
mean that this can be done retrospectively once the derivatives of $N$ have been  
caculated (see below).

As discussed in Section~\ref{sec:deltaN}, we need to calculate the amount of 
expansion between a flat and an equal-energy density hypersurface, $N_*^f(\phi_*,\chi_*)$, 
which we obtain by solving Eqs.~(\ref{equ:eoms}) for different initial values $\phi_*$ and $\chi_*$. 
The final hypersurface needs to be chosen in such a way that the scalar fields have
decayed into radiation, so that the curvature perturbation is conserved.
To find a suitable final energy density or, equivalently, final Hubble rate, we
first run the code from the initial condition $\phi_i=\phi_{\rm crit}$ and a suitable 
value of $\chi_i$. We follow the evolution until $99.9 \%$ of the energy density was
in the radiation component, and use the corresponding Hubble rate $H_f$ to define the 
final hypersurface. We also record the values
$\phi_*$ and $\chi_*$, which the fields take 
$60$ e-folds before the end 
of inflation. The numerical system is then 
re-run from initial conditions which are slightly different 
to the recorded $\phi_*$ and $\chi_*$, and the 
number of e-folds to the same final value of $H_f$ of the Hubble rate 
recorded, in such a way as to build up a finite difference 
approximation for the derivatives of $N_*^f(\phi_*,\chi_*)$. 

Once we have obtained $N_*^f(\phi_*,\chi_*)$, we calculate the spectral 
index $\ns$ and $\fnl$ using Eqs.~(\ref{eq:spectrum}) and (\ref{eq6}), respectively.
The results in Fig.~\ref{conns} show that for all the parameters we considered, the 
spectral index is contained in the region 
$0.9649 \le \ns \le 0.9701$. Therefore they are all compatible with the observations.

On the other hand, Fig.~\ref{confnl} shows that a 
wide range of values for $\fnl$ are obtained,  
$7\times 10^{-3} \lesssim \fnl  \lesssim 5 \times 10^5$. 
All parameters giving $\fnl \gtrsim 100$ are excluded by 
WMAP measurements \cite{wmap}, constraining the parameter space.
For $\chi_i\lesssim 0.1v^2/\mpl$, values $v\lesssim 0.07\mpl$ are ruled out,
and larger values of $\chi_i$ are not constrained.

Furthermore, Fig.~\ref{confnl} confirms the validity  
of the analytical arguments and estimates in Section~\ref{sec:AG} . To a good approximation, we can identify $\chi_*\approx \chi_i$. 
Our results show that as long as $\chi_i$ is small enough, $\fnl$ is independent of it and decreases by one 
order of magnitude for a two orders of magnitude increase in 
$v/\mpl$, as predicted by Eq.~(\ref{anpred}).
Towards the right, at $\chi_i\sim 0.1v^2/\mpl$, $\fnl$ starts to fall in accordance with Eq.~(\ref{equ:fnltrans}), and reaches a constant value $\fnl \approx 0.007$ in good agreement with Eq.~(\ref{anpredphi}). 

As mentioned above in this section, with the parametrization given in Eq.~(\ref{equ:paramchoices}), we can fix the parameters of the model retrospectively, constrained by the amplitude of the power spectrum. As can be seen in Fig.~\ref{conLambda}, $10^{-6}\lesssim\lambda\lesssim10^{-25}$. 

An issue of importance is whether quantum fluctuations will 
affect which side of the potential the waterfall field is rolling along. 
This happens if the range of quantum fluctuations is greater 
than the distance of 
$\chi$ to its hilltop. It is problematic because such a scenario leads to 
a configuration with domain walls, which we want to avoid. 
For a massless scalar field, the range is given by $H_*/(2\pi)$. For a massive field, 
$H_*/(2\pi)$ is an upper bound. Thus, excluding parameters such that $\chi_*/H_* \le 1$ 
settles the issue. As Fig.~\ref{conDH} shows, the excluded values are approximately
$v \le 5 \times 10^{-3}\mpl$ and $\chi_i< v^2/\mpl $. 
\begin{figure}
\centering
\includegraphics[width= 8.6cm]{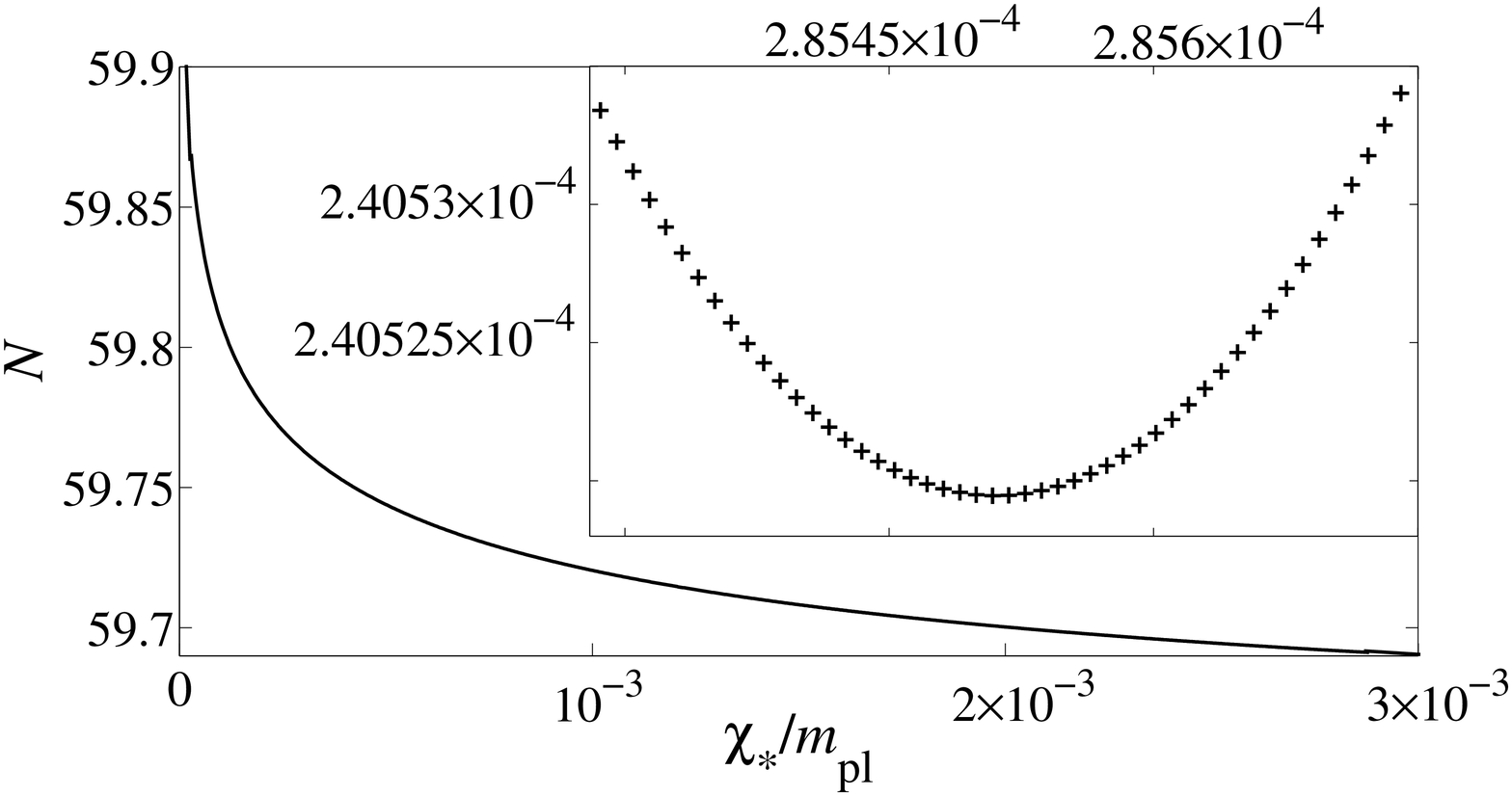}
\caption{$N$ as a function of $\chi_*$ for the parameters shown in Eq.~(\ref{equ:paramchoices}). The result in Section~\ref{sec:deltaN} assume that the quadratic Taylor expansion in Eq.~(\ref{eq2}) is a good approximation. This is clearly not the case for the whole range of $\chi_*$, but for a range that corresponds to the values present in one comoving volume corresponding to the currently observable universe it is. This is demonstrated by the inset, which shows the data for such a range with the linear term subtracted, together with a quadratic fit.}
\label{fit}
\end{figure}
\begin{figure}
\centering
\includegraphics[width= 8.6cm]{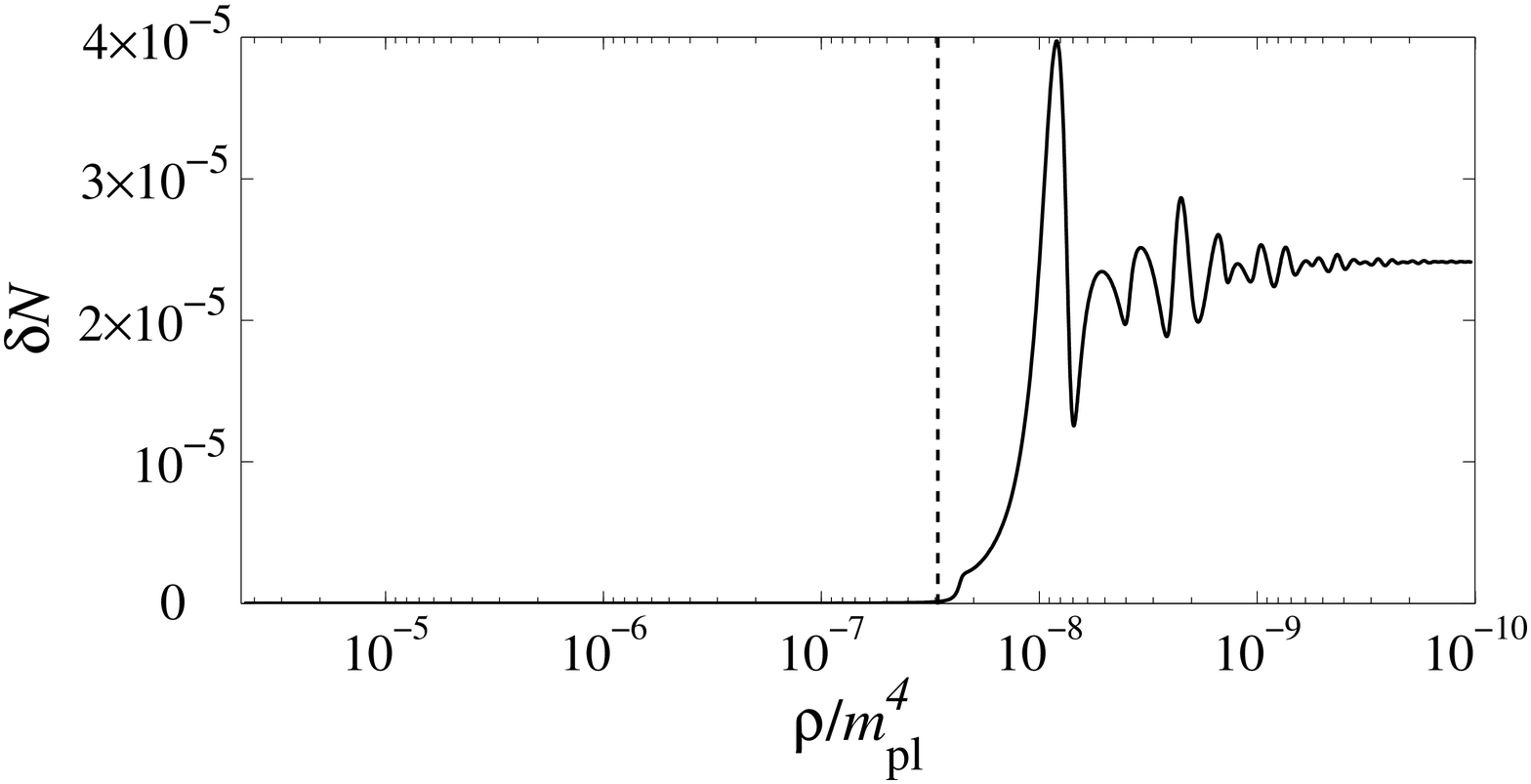}
\caption{Comparison of two initial conditions $\chi_i$ and $\chi_i+\delta\chi_i$ with parameters in Eq.~(\ref{equ:paramchoices}) and $\delta\chi_i=1.4\times 10^{-7}\mpl$. The plotted function is the difference in $N$ as a function of enery density $\rho$, and corresponds to the curvature perturbation between points with the two different initial conditions.
The vertical dashed line  indicates when the waterfall field $\chi$ starts rolling down the 
ridge.}
\label{dNevo}
\end{figure}
\section{A working example}
\label{sec:workExample}
Our results indicate that there is a fairly large window of parameters that are compatible with 
observations but lead to observable non-Gaussianity. We pick pick one representative example
for more detailed study:
\begin{eqnarray}
\label{eq:example}
v&=&0.2\mpl,\nonumber\\
\lambda &=& 5.322\times 10^{-15},\nonumber\\
g^2 &=& 2.128\times 10^{-19},\nonumber\\
\phi_i &=& \phi_{\rm crit},\nonumber\\
\chi_i &=& 10^{-3}v.
\end{eqnarray}
These parameters give
\begin{eqnarray}
\fnl&=&24.14,\nonumber\\
n_s&=&0.968.\nonumber\\
\end{eqnarray}
Besides a value of the spectral index in the centre of the 
permitted observational range, these parameters give us a large, but 
observationally compatible, value of $\fnl$. Indeed the value is close to the 
favoured WMAP value, $\fnl \approx 20 \pm 10 $.   

In Fig.~\ref{fit}, we show $N$ for a wide range of field values $\chi_*$. 
The expressions (\ref{eq6}) and (\ref{eq:spectrum}) for $\fnl$ and $n_s$ assume quadratic Taylor expansion (\ref{eq2}) of this function. 
It is clear from the figure that this assumption is not valid over the whole range shown. However, the actual range of $\chi_i$ present in a comoving volume corresponding to the universe observable today is smaller, roughly $\delta\chi_i\approx \sqrt{N/2\pi}H_*$, where $N\approx 60$. 
The inset shows this range of $\chi_*$, with the linear term subtracted. The remaining contribution is small and almost exactly quadratic, demonstrating that the Taylor expansion (\ref{eq2}) is valid.

It is interesting to confirm our expectation that the 
dominant contribution to $\delta N$ comes from the 
rolling of the waterfall field from its hilltop. 
This can be seen clearly in Fig.\ref{dNevo} which plots the evolution 
of $\delta N$ together with the point at which the waterfall field's 
evolution becomes significant.

\begin{figure}
\centering
\includegraphics[width= 8.6cm]{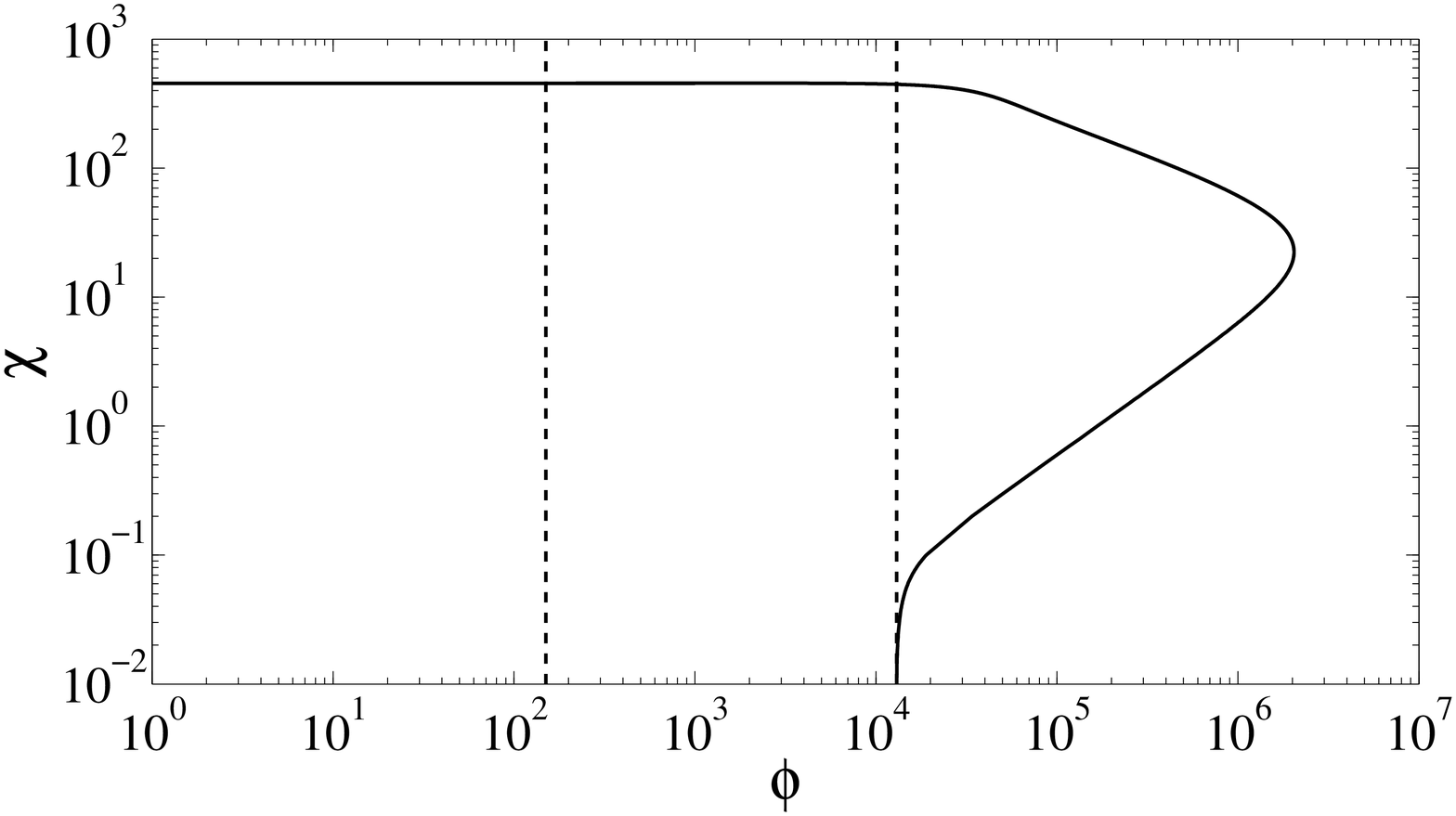}\caption{A plot of the boundary in field space on which quantum fluctuations become greater than the classical 
rolling of the 
fields. The three regions, described in the text, demarcate initial conditions on the surface for which $\chi$ does not reach $\chi=0$ before 
$\phi$ reaches $\phi_{\rm crit}$ (far left), initial conditions which do (middle), and initial conditions for which eternal inflation does not end 
(far right). Because of the symmetry of the potential, only positive values of $\phi$ and $\chi$ are shown.}
\label{fig:Eternal}
\end{figure}

Finally, we ask whether 
the initial conditions we have assumed in this section are reasonable. 
Importantly, for the parameter values given in example (\ref{eq:example}), 
we find that the condition, given in Section~\ref{sec:ics}, for mass structure to be correct
for $\chi$ to evolve classically to $\chi=0$, $\phi_{\rm e}^2>m^2/g^2$, is met if $\phi_{\rm e} > 32\mpl$. 
This is necessarily satisfied when $\phi$ is above $\phi_{\rm crit}$ in this 
case. Of course, in practice we will need $\phi$ to be significantly larger than 
this value in order that 
$\chi$ has time to evolve to zero. Numerically we find that $\phi_{\rm e}>150\mpl$ is acceptable 
for $\chi\ll v^2/\mpl$ at $\phi_{\rm crit}$, which 
is satisfied on a significant proportion of  the surface defined by Eq.~(\ref{eq:eternal}). This can be seen 
by considering Fig.~\ref{fig:Eternal}, which plots the surface on which eternal 
inflation ends for the parameter values at hand. For values of $\phi\gg150\mpl$ the evolution 
is predominantly in the $\chi$ direction. Since the surface is complicated, an evolution which leaves the eternal regime may 
re-enter it, as can be seen from the figure. The dashed lines demarcate three regimes, trajectories 
originating from the surface in the region on the far left do not evolve to $\chi=0$ 
before $\phi$ reaches $\phi_{\rm crit}$, while trajectories in the middle region do.  Trajectories in the region on the 
far right originating from the upper eternal inflation boundary evolve classically towards $\chi=0$, but 
reenter an eternal inflationary regime. The vast majority of trajectories which remain classical, however, 
evolve to $\chi \ll v^2/\mpl$ when $\phi = \phi_{\rm crit}$.

\section{Conclusions}

In this paper we have
demonstrated
that inflation sourced by the hybrid potential Eq.~(\ref{eq:hybridPot}) 
can give rise to an observationally significant non-Gaussianity while satisfying WMAP 
constraints on the power spectrum. 
The conditions required for this are
summarised in Eq.~(\ref{eq:ourParameters}). 
This is an example of the 
hilltop mechanism for producing non-Gaussianity first identified for axion models in 
Ref.~\cite{Kim:2010ud}.

In contrast with original hybrid inflation, where the waterfall field is heavy 
and cannot contribute significantly to the curvature perturbation on large scales \cite{heavy}, 
we take the waterfall field to be light. 
Our scenario also
leads to more than $60$ e-folds of accelerated expansion 
after the waterfall transition. This avoids issues of primordial black holes and 
domain walls. While earlier work has considered  
the hybrid model with a light waterfall field and the resulting non-Gaussianity 
\cite{barnaby, Abolhasani:2011yp} (and hybrid inflation 
with two 
light inflaton fields \cite{largeNG1,inhom2}), 
we believe that our work
is the only study of non-Gaussianity in the case in which 
more than $60$ e-folds of evolution occurs after the hybrid transition. 

Numerically we have only exhaustively studied the 
range of parameters discussed in Section~\ref{sec:numerical}, but we have also probed 
several other parameter choices which lead to $60$ e-folds after the hybrid 
transition, including a number which  
are outside of the regime summarised by Eq.~(\ref{eq:ourParameters}). 
Inside this regime analytic expectations are confirmed remarkably well by our simulations, 
which suggests that we should be able to trust them even outside it as long as the assumptions thay are based on are satisfied.
Outside of this regime
our non-exhaustive probing has not found any 
parameter choices which lead to a significant non-Gaussianity. It would be 
interesting, however, to probe an extended parameter range more carefully in order to 
determine whether 
the conditions Eq.~(\ref{eq:ourParameters}) are the only ones which lead to a signal.

Finally we note that as discussed in Section~\ref{sec:ics} an interesting 
feature of the hybrid model is that it offers an 
explanation for the initial conditions at horizon crossing which lead to 
a large non-Gaussianity. While parameter choices are necessarily fine tuned, therefore, 
the initial conditions need not be, at least for some subset of the parameters 
which can produce a large non-Gaussianity. This is in contrast to other two field 
models which produce a large non-Gaussianity due to inflationary dynamics \cite{largeNG1, largeNG2,Elliston:2011dr}, which 
require both carefully choosen model parameters, and offer no explanation for the origin of the
finely tuned horizon crossing conditions.

\section*{Acknowledgements}
We thank David Seery for a careful reading of the manuscript and for useful comments. 
AR and DJM were supported by STFC, and SO by EPSRC. The work was also partly supported by the Royal Society.


\end{document}